\documentclass[prd,aps,tightenlines,preprint,nofootinbib,a4paper]{revtex4}
\usepackage{epsfig}
\usepackage{amsmath}
\usepackage{appendix}

\begin{document}

\title{Hermitian matrix diagonalization and its symmetry properties}

\author{S. H. Chiu\footnote{schiu@mail.cgu.edu.tw}}
\affiliation{Physics Group, CGE, Chang Gung University, 
Taoyuan 33302, Taiwan}

\author{T. K. Kuo\footnote{tkkuo@purdue.edu}}
\affiliation{Department of Physics, Purdue University, West Lafayette, IN 47907, USA}


\begin{abstract}
A hermitian matrix can be parametrized by a set consisting of its determinant and
the eigenvalues of its submatrices.
We established a group of equations which connect these variables with the mixing parameters
of diagonalization. These equations are simple in structure and manifestly invariant in form
under the symmetry operations of dilatation, translation, rephasing and permutation.
When applied to the problem of neutrino oscillation in matter they produced 
two new ``matter invariants" which are confirmed by available data.

\end{abstract}


\maketitle

\pagenumbering{arabic}




\section{Introduction}

The study of hermitian matrices has a long history.
Their diagonalization, in particular, has been used in a wide variety of topics.
It is also known to be invariant under a number of symmetry operations, 
viz., dilatation, translation, rephasing, and permutation. However, some of these symmetries
seem to have been treated with benign neglect.
For instance, rephasing invariance is used to fixed certain phases (somewhat arbitrarily), 
as in the Particle Data Group (PDG) \cite{PDG} flavor mixing parameters, but not imposed generally. 
In the same vein,  permutation symmetry allows one to adopt a convention in ordering the
states, usually in ascending order of their eigenvalues, thereby obscuring the underlying
freedom to rearrange them arbitrarily.

In this paper we will formulate diagonalization so that all of its relations are manifestly invariant 
in form under the afore-mentioned symmetries. To do that, we need to use parameters which transform simply.
Thus, we propose to parametrize a matrix $(M)$, not by its matrix elements, but by combinations composed
of its eigenvalues and those of its submatrices, $M_{\alpha}$, which are obtained by deleting the $\alpha-$th
row and column from $M$. We are then able to find a set of equations connecting these parameters
with the mixing parameters $(W_{\alpha i}=|V_{\alpha i}|^{2})$. These equations are manifestly invariant
in form under the symmetry operations.
They are also very simple and intuitive. As an application, we studied the problem of neutrino oscillation in matter.
We found that these equations led to two new ``matter invariants", similar to the well-known ones 
involving the Jarlskog CP-violating parameter \cite{J}.

\section{Parametrization of Hermitian Matrices}

It is well-known that the fundamentals of hermitian matrix diagonalization are governed by the characteristic
polynomials,
\begin{equation}
\det (\lambda I-M)=\det (\lambda I-M_{D})=\Pi (\lambda-\lambda_{i}).
\end{equation}
Here, $M_{D}$ refers to the diagonal matrix with eigenvalues $\lambda_{i}$. From this one infers that
diagonalization is invariant in form under a number of symmetry operations:
1) Dilatation, 2) Translation, 3) Rephasing, and 4) Permutation.

To fully appreciate the implications of these symmetries, 
it is important to use parameters that transform simply under them.
Of the three groups of relevant variables $(M_{\alpha \beta};\lambda_{i};V_{\alpha i})$,
a natural choice for the mixing parameters seems to be $W_{\alpha i}=|V_{\alpha i}|^{2}$,
which are invariant under dilatation, translation, and rephasing, and transform as tensors
under permutation. However, some regrouping is needed for $M_{\alpha \beta}$ and 
$\lambda_{i}$.

We begin by using the Dirac notation to write a hermitian operator $M$ (also used to denote
the matrix) in terms of $|\alpha \rangle$, the flavor basis ket, and $|i\rangle$, the eigenvalues basis vector, 
\begin{equation}
M=\Sigma M_{\alpha \beta} |\alpha \rangle \langle \beta|=\Sigma \lambda_{i}|i\rangle \langle i|.
\end{equation}
Thus, the matrix elements are given by 
\begin{equation}
M_{\alpha \beta}=\langle \alpha |M|\beta \rangle,
\end{equation}
\begin{equation}
M_{ii}=\langle i|M_{D}|i\rangle,
\end{equation}
where $M_{D}$ = diagonal matrix. Note that the above definition is manifestly independent
of the ordering of $| \alpha \rangle$ or $|i \rangle$, corresponding to the permutation symmetry
$(S_{n} )_{F}\times (S_{n})_{E}$.

To proceed, in order to make things simple and concrete, we will first study the diagonalization
of $3\times 3$ matrices. We will also use the notation of leptonic flavor physics, so that 
$\alpha=(e,\mu,\tau)$ and $i=(1,2,3)$.

Now, the state $|\alpha\rangle$  can undergo a rephasing transformation
\begin{equation}
|\alpha\rangle \rightarrow e^{-i\phi_{\alpha}}|\alpha \rangle.
\end{equation}
Correspondingly, for a matrix element,
\begin{equation}
M_{\alpha \beta}=e^{i\phi_{\alpha \beta}}|M_{\alpha \beta}|, \, \, \,  \alpha \neq \beta, 
\end{equation}
\begin{equation}
\phi_{\alpha \beta} \rightarrow \phi_{\alpha \beta}+(\phi_{\alpha}-\phi_{\beta}).
\end{equation}
Thus, of the three phases in $\phi_{\alpha \beta}$, rephasing can remove two, leaving
a rephasing invariant phase
\begin{equation}
\phi=\phi_{\alpha \beta}+\phi_{\beta \gamma}+\phi_{\gamma \alpha}.
\end{equation}
It follows that the rephasing invariant parameters of $M$ consist of $M_{\alpha \alpha}$,
$|M_{\alpha \beta}| \, \, \, (\alpha \neq \beta)$, and a phase, which can be taken as $\phi$,
giving altogether $3^{2}-2=7$ parameters.

However, they do not transform simply under the symmetry operations. 
To remedy this we propose to use a set which is composed of eigenvalues
of $M$ and $M_{\alpha}$, the submatrix of $M$ with the $\alpha-$th row and column deleted.
Specifically, we have the set
$(\det M, d_{\alpha},t_{\alpha})$, $\alpha=(e,\mu,\tau)$, with
\begin{equation}\label{da}
d_{\alpha}=\det M_{\alpha}=\xi^{(\alpha)}_{\rho_{1}} \cdot \xi^{(\alpha)}_{\rho_{2}} ,
\end{equation}
\begin{equation}\label{ta}
t_{\alpha}=tr M_{\alpha}= \xi^{(\alpha)}_{\rho_{1}} + \xi^{(\alpha)}_{\rho_{2}} ,
\end{equation}
where $\xi^{(\alpha)}_{\rho_{i}}$, $i=1,2$, are the eigenvalues of $M_{\alpha}$.
All of these variables transform simply under the symmetry operations:

\begin{enumerate}

\item Dilatation: $M\rightarrow rM$, $\det M \rightarrow r^{3}(\det M)$, 
$d_{\alpha} \rightarrow r^{2}d_{\alpha}$, $t_{\alpha} \rightarrow r t_{\alpha}$.

\item Translation: $M \rightarrow M+(\Delta \lambda) I$, 
$d_{\alpha} \rightarrow d_{\alpha}+(\Delta \lambda) t_{\alpha}$, $t_{\alpha} \rightarrow t_{\alpha}+2(\Delta\lambda)$.

\item Rephasing invariance is satisfied by all.

\item Under $S_{3}$, the parameters transform as tensors. Note that, under $S_{2}$ (for $M_{\alpha}$),
they are invariant.
\end{enumerate}

Given $(\det M, d_{\alpha}, t_{\alpha})$, we can also solve for the matrix elements $M_{\alpha \beta}$
and $\phi$. The diagonal elements can be determined by $t_{\alpha}$. 
The off-diagonal ones $|M_{\alpha \beta}|^{2}$, $\alpha \neq \beta$, are given by $d_{\alpha}$.
And, finally, $\phi$ is fixed by $\det M$.

Thus, the set $(\det M, d_{\alpha}, t_{\alpha})$, $\alpha=(e,\mu,\tau)$, which transform simply under symmetry
operations, is suitable as a parametrization of the rephasing invariant part of a $3\times 3$ hermitian matrix.


This parametrization can be generalized to $n \times n$ hermitian matrices. Let us define
\begin{equation}\label{nbyn}
 \Pi_{i=1}^{n-1}(\lambda-\xi_{\rho_{i}}^{(\alpha)})=\Sigma (-1)^{m}\lambda^{n-1-m}d_{\alpha}^{(m)},
\end{equation}
with
\begin{equation}
d_{\alpha}^{(0)}=1,
\end{equation}
\begin{equation}
d_{\alpha}^{(n-1)}=\det M_{\alpha}.
\end{equation}
Here, $\xi_{\rho_{i}}^{(\alpha)}$, $\rho_{i}=(\beta, \gamma, ...)$, denote the $(n-1)$ eigenvalues of $M_{\alpha}$,
so that $d_{\alpha}^{(m)}$ is the symmetric sum of the products of $m$ non-repeating eigenvalues of $M_{\alpha}$.
$d_{\alpha}^{(m)}$ can also be expressed in terms of $M_{\alpha \beta}$ by expanding $\det(\lambda I_{(n-1)}-M_{\alpha})$
directly.

We now propose to use the set $(\det M, d_{\alpha}^{(m)})$, $m=(1, ....., n-1)$, to parametrize an $n \times n$
hermitian matrix. Some notable properties of the set are:
\begin{enumerate}
\item The number of variables in the set is $[1+n(n-1)]$, 
which is the same as that of the rephasing invariant part of $M$, given by $[n^{2}-(n-1)]$, where
$(n-1)$ phases are removed by rephasing. Note also that, diagonalization produces $n$ eigenvalues, with
the help of $(n-1)^{2}$ rephasing invariant parameters $(W_{\alpha i})$, yielding again the same total $[n+(n-1)^{2}]$.
\item Dilatation: $M\rightarrow r M$, $d_{\alpha}^{(m)} \rightarrow r^{m} d_{\alpha}^{(m)}$.
\item Translation: $M \rightarrow M+(\Delta \lambda) I$, all eigenvalues are boosted by the same $\Delta \lambda$, and
$d_{\alpha}^{(m)} \rightarrow d_{\alpha}^{(m)}+(n-m)(\Delta \lambda)d_{\alpha}^{(m-1)}$. 
\item Rephasing invariance is satisfied, together with $W_{\alpha i}$.
\item All variables transform as tensors under $S_{n}$. They are also invariant under $S_{n-1}^{(\alpha)}$,
for permutations pertaining to $M_{\alpha}$.

\end{enumerate}

Thus, the set, $(\det M, d_{\alpha}^{(m)})$ plus $W_{\alpha i}$, seems well-suited for diagonalization.
Note that the use of eigenvalues, which transform simply under all symmetries, enable the parameters
to also transform simply.
In Sec. V, we will use this parametrization to establish a set of simple equations which connect
the flavor variables with the mixing parameters and the eigenvalues of the matrix.

\section{$3 \times 3$ matrices}

We will now proceed to tackle the problem of the diagonalization of a $3\times 3$ matrix.
Following the previous section, we use the parameter set for $M$:
$(\det M,d_{\alpha},t_{\alpha})$, $\alpha=(e,\mu,\tau)$, with $d_{\alpha}=\det M_{\alpha}$ and
$t_{\alpha}=tr M_{\alpha}$, as in Eqs.(\ref{da}) and (\ref{ta}).   
Correspondingly, for the diagonal matrix $M_{D}$, we have $(\det M_{D},d_{i},t_{i})$, $i=(1,2,3)$,
\begin{equation}
d_{i}=\lambda_{j}\lambda_{k},
\end{equation}
\begin{equation}
t_{i}=\lambda_{j}+\lambda_{k},
\end{equation}
with $(i,j,k)$=cyclic permutation of $(1,2,3)$.

Consider the expansion of the characteristic polynomial
\begin{equation}
\det (\lambda I-M)=\lambda^{3}-\lambda^{2}(\frac{1}{2}\Sigma t_{\alpha})+\lambda (\Sigma d_{\alpha})-\det M,
\end{equation}
and that of $M_{D}$,
\begin{equation}
\det (\lambda I-M_{D})=\lambda^{3}-\lambda^{2}(\frac{1}{2}\Sigma t_{i})+\lambda (\Sigma d_{i})-\det M_{D}.
\end{equation}
Since they are equal to each other for arbitrary $\lambda$, we find
\begin{equation}
\det M=\det M_{D},
\end{equation}
\begin{equation}
\Sigma d_{\alpha}=\Sigma d_{i},
\end{equation}
\begin{equation}
\Sigma t_{\alpha}=\Sigma t_{i}.
\end{equation}

Now we turn to the case of a perturbation on $M$,   
\begin{equation}
M \rightarrow M + \varepsilon |\alpha \rangle \langle \alpha |.
\end{equation}
To first order in $\varepsilon$, the eigenvalues change by $\langle i|\Delta M |i\rangle$, i.e., 
\begin{equation}
\lambda_{i} \rightarrow \lambda_{i}+\varepsilon \langle i|\alpha \rangle \langle \alpha |i \rangle
=\lambda_{i}+\varepsilon \cdot W_{\alpha i}.
\end{equation}
Also, 
\begin{equation}\label{23}
\det[\lambda I_{3}-(M+\varepsilon |\alpha \rangle \langle \alpha|)]=\det (\lambda I_{3}-M)-
\varepsilon \cdot \det (\lambda I_{2}-M_{\alpha}),
\end{equation}
where $I_{3}$  $(I_{2})$ denote the $ 3\times 3$ $(2 \times 2)$ identity matrix. From Eq.(\ref{nbyn}), we have 
\begin{eqnarray}\label{A}
\det(\lambda I_{2}-M_{\alpha})&=&\Pi (\lambda -\xi_{\rho_{i}}^{(\alpha)}) \nonumber \\
&=&\Sigma_{m=0}^{2}(-1)^{m} \lambda^{2-m}d_{\alpha}^{(m)} \nonumber \\
&=&\lambda^{2}-\lambda(t_{\alpha})+d_{\alpha}.
\end{eqnarray}
Also,
\begin{eqnarray}\label{B}
\det [\lambda I_{3}-(M+\varepsilon|\alpha \rangle \langle \alpha |)]
&=&\Pi(\lambda -\lambda_{i}-\varepsilon \cdot W_{\alpha i}) \nonumber \\
&\cong& \Pi (\lambda -\lambda_{i})-\varepsilon \cdot \Sigma [W_{\alpha i}\cdot \Pi_{i \neq j}'(\lambda-\lambda_{j})].
\end{eqnarray}
Equating the coefficients of $\varepsilon \lambda^{m}$ of Eqs.(\ref{23}-\ref{B}), we have
\begin{equation}\label{dalpha}
d_{\alpha}=\Sigma W_{\alpha i}d_{i},
\end{equation}
\begin{equation}\label{talpha}
t_{\alpha}=\Sigma W_{\alpha i} t_{i}.
\end{equation}
Together with
\begin{equation}\label{detm}
\det M=\Pi \lambda_{i},
\end{equation}
these seven equations completely specify the diagonalization of a $3\times 3$ hermitian matrix.
(Note that, the equations $\Sigma d_{\alpha}=\Sigma d_{i}$ and $\Sigma t_{\alpha}=\Sigma t_{i}$ are
given by summing these equations.)

Eqs.(\ref{dalpha}), (\ref{talpha}), and (\ref{detm}) are the main results of this paper,
and a direct proof of the three equations is given in Appendix A.
They relate the parameters in the
flavor space ($\det M$,$d_{\alpha}$,$t_{\alpha}$) with those of the eigenvalue space ($\lambda_{i}$) and
the mixing parameters ($W_{\alpha i}$). Some of their salient features are: 
\begin{enumerate}
\item They are manifestly invariant under dilatation and rephasing. 
They are covariant under $(S_{3})_{F} \times (S_{3})_{E}$, and invariant under permutation in $M_{\alpha}$
and $M_{i}$, $(S_{2} \times S_{2})$. As for translation, for which $\Delta (\det M)=(\Delta \lambda)(\Sigma d_{\alpha})$,
$\Delta d_{\alpha}=(\Delta \lambda)t_{\alpha}$, $\Delta t_{\alpha}=2(\Delta \lambda)$,
these equations are ``knit" together in succession, and the whole set is invariant. \\
\item We can solve for $W_{\alpha i}$ by taking the sum $\lambda^{2}-\lambda t_{\alpha}+d_{\alpha}=
\Sigma W_{\alpha i}(\lambda^{2}-\lambda t_{i}+d_{i})$, from which, if we take $\lambda=\lambda_{i}$, only
$W_{\alpha i}$ survives and 
\begin{equation}
W_{\alpha i}=\frac{(\lambda_{i}-\xi_{\rho_{1}}^{(\alpha)})(\lambda_{i}-\xi_{\rho_{2}}^{(\alpha)})}{(\lambda_{i}-\lambda_{j})(\lambda_{i}-\lambda_{k})}
\end{equation}
which is the eigenvector-eigenvalue identities given earlier \cite{Denton1,Denton2}.  
(In Ref.\cite{Denton1}, a similar proof was also given in the context of first order perturbation theory.) 
Thus, these two approaches are closely related to each other.
However, we believe that these equations are more transparent in appearance and easier
to grasp.  Also, since each equations contains only a subset of parameters in the problem,
they may be more adaptable for applications in physics, where experimental data are often uneven,
favoring part of the parameter set over others. An example will be given in the next section. 
\item We can also consider the ``inverse" problem, expressing $d_{i}^{(m)}$ in terms of the others.
To do this we may turn to $w_{\alpha i}$ \cite{KuoLee}, the cofactor of $W_{\alpha i}$, which satisfies   
\begin{equation}
w^{T}W=Ww^{T}=\det W \equiv D,
\end{equation}
\begin{equation}
D=\Sigma _{\alpha}w_{\alpha i}=\Sigma_{i} w_{\alpha i}.
\end{equation}
Thus, we have
\begin{equation}
d_{i}=\frac{1}{D}(w^{T})_{i \alpha}d_{\alpha},
\end{equation}
\begin{equation}
t_{i}=\frac{1}{D}(w^{T})_{i \alpha}t_{\alpha}.
\end{equation}
Note also that, in terms of the notation adopted earlier, we have  
\begin{equation}
\widetilde{t}_{i}=t_{j}-t_{k}=-\widetilde{\lambda}_{i},
\end{equation}
\begin{equation}
\widetilde{d}_{i}=d_{j}-d_{k}=-\widetilde{\lambda}_{i}\lambda_{i},
\end{equation}
\begin{equation}
\Sigma d_{i}\widetilde{t}_{i}=-\Sigma t_{i}\widetilde{d}_{i}=\Pi \widetilde{\lambda}_{l}.
\end{equation}
In stead of the formula $D=(\Sigma d_{\alpha}\widetilde{t}_{\alpha})/\Pi \widetilde{\lambda}_{l}$ 
given in Ref.\cite{ck2022}, 
we now have a more suggestive expression
\begin{equation}
D=(\Sigma d_{\alpha} \widetilde{t}_{\alpha})/(\Sigma d_{i} \widetilde{t}_{i}),
\end{equation}
which vividly exhibits the connection between mixing, flavor, and eigenvalues in diagonalization. 
\item As usual, it is instructive to consider $2 \times 2$ matrices. In this case, the parameter set is
($\det M,t_{\alpha}$), $\alpha=(e,\mu)$, $(t_{e}=M_{\mu \mu},t_{\mu}=M_{ee})$.
The relevant equations are $\det M=\det M_{D}$, 
$t_{\alpha}=\Sigma W_{\alpha i} t_{i}$, ($i=1,2$), $t_{1}=\lambda_{2}$, $t_{2}=\lambda_{1}$, or
\begin{equation}\label{mLL}
\det M=\lambda_{1}\lambda_{2},
\end{equation}
\begin{equation}\label{teWW}
t_{e}=W_{e1}\lambda_{2}+W_{e2}\lambda_{1},
\end{equation}
\begin{equation}\label{tmuWW}
t_{\mu}=W_{\mu 1}\lambda_{2}+W_{\mu 2}\lambda_{1}.
\end{equation}
Eq.(\ref{mLL}) plus the sum of Eqs.(\ref{teWW}) and (\ref{tmuWW}) gives $\det M=\lambda_{1}\lambda_{2}$,
$M_{ee}+M_{\mu \mu}=\lambda_{1}+\lambda_{2}$, which are the familiar formulae to determine $\lambda_{i}$.
Subtracting the two equations yield
\begin{equation}\label{WWD}
W_{e1}-W_{\mu 1}=D=(t_{e}-t_{\mu})/(\lambda_{2}-\lambda_{1}),
\end{equation}
where, in the usual notation, $W_{e1}=\cos^{2}\theta$, $D=\cos 2\theta$.
Eq.(\ref{WWD}) is just Eq.(42) of Ref.\cite{ck2022}.     
\end{enumerate}


 \begin{figure}
\caption{$I_{d}$ and $I_{t}$ as functions of $A$. Here, $\lambda_{i}$ in $I_{d}$ and $I_{t}$ are taken 
as the eigenvalues of the squared effective neutrino mass matrix 
($\lambda_{i}=m^{2}_{i}$). We adopt the values of $W_{\alpha i}$ in vacuum 
as: $W_{e1}=0.678$, $W_{e2}=0.301$, $W_{e3}=0.021$. With the use of $\delta_{0} \equiv m_{2}^{2}-m_{1}^{2}$ 
and $m_{3}^{2}-m_{2}^{2}=33 \delta_{0}$, we also assume $m_{1}^{2}=0.01\delta_{0}$ in vacuum,
but it does not affect the test of invariance. 
They are used to find approximate values of $W_{ei}(A)$ and $\lambda_{i}(A)$.} 
\centerline{\epsfig{file=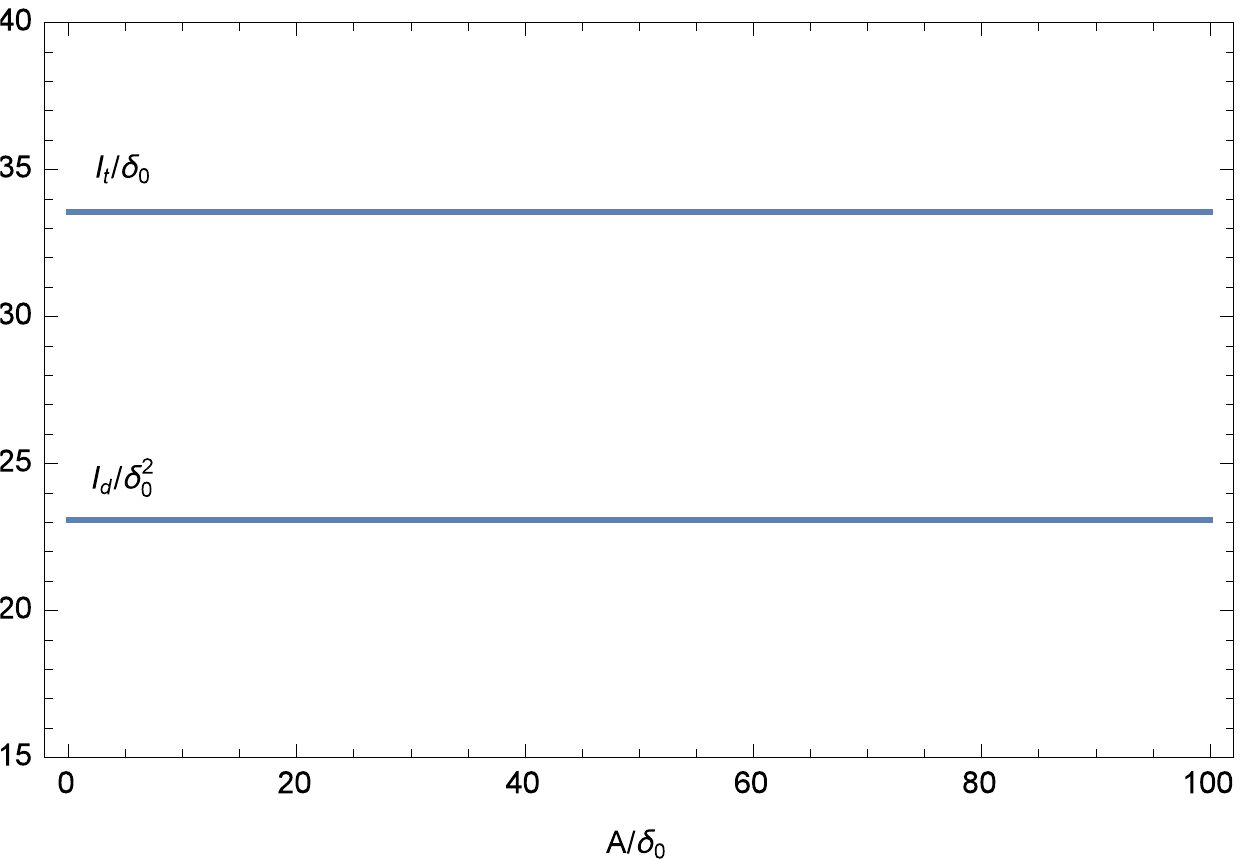,width= 10cm}}
\end{figure}

\section{Neutrino Oscillation and matter Invariants}

When a neutrino propagates through a medium, it picks up an effective mass.
Thus, the presence of matter calls for an additional term in the neutrino mass matrix \cite{Wolf,MS} 
\begin{equation}
M \rightarrow M(A)=M+A |e\rangle \langle e|,
\end{equation}
where $A=2\sqrt{2}G_{F}n_{e}E$. Diagonalizing $M(A)$ then yields the neutrino parameters in matter.
In the set of Eqs.(\ref{dalpha}), (\ref{talpha}), and (\ref{detm}), now considered as functions of $A$,
we can isolate the part related to $|e \rangle$,
\begin{equation}\label{Wdd}
\Sigma W_{ei} d_{i}=d_{e},
\end{equation}
\begin{equation}\label{Wtt}
\Sigma  W_{ei} t_{i}=t_{e}.
\end{equation}
Since $d_{e}$ and $t_{e}$ are obviously not affected by $A$,
the left-hand-side of Eqs.(\ref{Wdd}) and (\ref{Wtt}) must also be independent of $A$.
They are thus ``matter invariants", which can be explicitly written as
\begin{equation}\label{Id}
I_{d}=\Sigma ' W_{ei}\lambda_{j}\lambda_{k}.
\end{equation}
\begin{equation}\label{It}
I_{t}=\Sigma ' W_{ei}(\lambda_{j}+\lambda_{k}),
\end{equation}
where $\Sigma '$ indicates cyclic summation over the indices $(i,j,k)$.
They are similar to the well-known matter invariants $[J\cdot \Pi \widetilde{\lambda}_{l}]$,
$[J^{2}/\Pi W_{ei}]$, and the combination $[\Pi W_{ei}/(\Pi \widetilde{\lambda}_{l})^{2}]$ 
\cite{Nau,Tosh,Kimura}. 
Also, they are generalizations of the two flavor invariant: $t_{e}=W_{e1}\lambda_{2}+W_{e2}\lambda_{1}$.

Physically, an interesting feature of $I_{d}$ and $I_{t}$ is that they contain variables that
are better known experimentally. In fact, their vacuum values $(A=0)$ have been determined rather accurately.
More $A\neq 0$ data should also be available in the near future from the Long Baseline Experiments.  
Before that happens, there are theoretical calculations for the $A\neq 0$ data, 
such as the results obtained from the use of differential equations. It was found \cite{CK2018} that 
\begin{equation}\label{D1}
\frac{d}{dA}\lambda_{i}=W_{ei},
\end{equation}
\begin{equation}\label{D2}
\frac{1}{2}\frac{d}{dA}W_{ei}=W_{ei}[\frac{W_{ej}}{\lambda_{i}-\lambda_{j}}+\frac{W_{ek}}{\lambda_{i}-\lambda_{k}}],
\end{equation}
where all variables are $A$-dependent.

First of all, these equations can be used to verify the constancy of $I_{d}$ and $I_{t}$. Thus, using
\begin{equation}
\Sigma ' W_{ei}[\frac{W_{ej}}{\lambda_{i}-\lambda_{j}}+\frac{W_{ek}}{\lambda_{i}-\lambda_{k}}]\lambda_{j}\lambda_{k}
=-\Sigma '(W_{ei}W_{ej}\lambda_{k}),
\end{equation}
we can immediately verify that 
\begin{equation}
\frac{d}{dA}I_{d}=0,
\end{equation}
and, similarly,
\begin{equation}
\frac{d}{dA}I_{t}=0,
\end{equation}
showing that $(I_{d},I_{t})$ are indeed matter invariants. 
Eqs.(\ref{D1}), (\ref{D2}) were also used to obtain approximate solutions for $(W_{ei},\lambda_{i})$ \cite{CK2018}.  
We used these results to test $I_{d}$ and $I_{t}$ in Fig.1. 
It is seen that both $I_{d}$ and $I_{t}$ are very nearly constant over the whole range of $A$,
even though the individual parameters vary considerably, as shown in Figs. 1 and 2 of Ref. \cite{CK2018}.   
This can be taken to mean that the invariance of $I_{d}$ and $I_{t}$ is verified. 
At the same time, we can also infer that the approximation used in Ref. \cite{CK2018} is robust.  
Of course, it would be most interesting if measurements of $(W_{ei},\lambda_{i})$
become available, and direct tests of $I_{d}$ and $I_{t}$ can be carried out.

The invariance of $I_{d}$ and $I_{t}$ illustrates  a general feature of theories which support direct relations between $M_{\alpha \beta}$ and $(\lambda_{i},W_{\alpha i})$.
While $(\lambda_{i},W_{\alpha i})$ are experimentally accessible, $M_{\alpha \beta}$ are usually theoretical constructs.
These relations can transform properties of $M_{\alpha \beta}$ into constraints on $(\lambda_{i},W_{\alpha i})$.
Another such example is the well-known matter invariant, $J\cdot \Pi (\lambda_{i}-\lambda_{j})$,
deriving \cite{HaSc} from the relation,    
$Im[M_{\alpha \beta}M_{\beta \gamma}M_{\gamma \alpha}]=J\cdot \Pi(\lambda_{i}-\lambda_{j})$.
The behavior of this relation under $(S_{3})_{F} \times (S_{3})_{E}$ is also noteworthy.
While $J \sim ({\tilde{\bf1}},{\tilde{\bf1}})$, $\Pi(\lambda_{i}-\lambda_{j}) \sim ({\bf1},{\tilde{\bf1}})$,
$M_{\alpha \beta} \sim (\bf{3} \times \bf{3}, \bf{1})$, 
$Im(M_{\alpha \beta} M_{\beta \gamma}M_{\gamma \alpha}) \sim (\tilde{\bf1},\bf{1})$,
since $(M_{\alpha \beta} M_{\beta \gamma}M_{\gamma \alpha}) \rightarrow 
(M_{\alpha \beta} M_{\beta \gamma}M_{\gamma \alpha})^{*}$ under any exchange $(\alpha,\beta)$.
Lastly, similar analyses may be applied to the renormalization of the mass matrices of
quarks and leptons, which we hope to present in the future.


\section{$n \times n$ matrices}

The generalization from $3 \times 3$ to $n \times n$ matrices is straightforward.
We will now outline the results.

First of all, for an $n \times n$ hermitian matrix, we may use the parameters 
$(\det M,d_{\alpha}^{(m)})$, $m=1, ...,(n-1)$, $\alpha =(\alpha_{1}, ..., \alpha_{n}).$
As before, $d_{\alpha}^{(m)}$ are defined by the formula
\begin{equation}
\det(\lambda I_{(n-1)}-M_{\alpha})=\Sigma_{m=0}^{n-1}(-1)^{m}\lambda^{n-1-m}d_{\alpha}^{(m)},
\end{equation}
with $d_{\alpha}^{(0)}=1$, $d_{\alpha}^{(n-1)}=\det M_{\alpha}$.
Correspondingly, for the diagonal matrix $M_{D}$ and its submatrices $M_{i}$, we have 
\begin{equation}
\det (\lambda I_{n-1)}-M_{i})=\Sigma_{m=0}^{n-1}(-1)^{m}\lambda^{n-1-m}d_{i}^{(m)},
\end{equation}
\begin{equation}
d_{i}^{(0)}=1,  \;  \; \; d_{i}^{(n-1)}=\det M_{i}.
\end{equation}
Here, $(\det M_{D},d_{i}^{(m)})$ is a redundant set, containing only $n$ independent parameters.

Following the same perturbation analysis as in Eq.(\ref{B}), except that now $\alpha=(\alpha_{1}, ..., \alpha_{n})$
and $i=1, ...,n$, we find the following set of equations
\begin{eqnarray}\label{V}
\det M &=&\det M_{D}, \; \; \; d_{\alpha}^{(m)} = \Sigma W_{\alpha i}d_{i}^{(m)}, \nonumber \\
m &=&(1, ... , n-1), \; \; i=(1, ... , n) \; \; \alpha =(\alpha_{1}, ... , \alpha_{n}).
\end{eqnarray}

As before, this set of equations are invariant in form under dilatation, translation, rephasing, and
permutation (covariant under $S_{n} \times S_{n}$ for $M$,  but invariant under 
$S_{n-1}\times S_{n-1}$ for $M_{\alpha}$.)
Also, if we sum up these equations
\begin{equation}
\Sigma (-1)^{m}\lambda^{n-1-m}d_{\alpha}^{(m)}=\Sigma W_{\alpha i}(\Sigma(-1)^{m}\lambda^{n-1-m}d_{i}^{(m)}),
\end{equation}
and choosing $\lambda=\lambda_{I}$, the only term that survives in $\Sigma_{i}$ is the $i=I$ term.
Thus, this would  extract the $W_{\alpha I}$ term, 
and $\Sigma (-1)^{m}\lambda_{I}^{n-1-m}d_{\alpha}^{(m)}=W_{\alpha I}[\Sigma (-1)^{m}\lambda_{I}^{n-1-m}d_{I}^{(m)}]$,
which is just the result given by the eigenvector-eigenvalue identities \cite{Denton1,Denton2}.
We may compare this to the traditional method for diagonalization, where one first find the
eigenvalues for $M$, then solve for the eigenvector, which are collected into the mixing matrix.
Here, one would first solve for the eigenvalues of $M$, and determine $d_{\alpha}^{(m)}$
from$M_{\alpha}$. 
After which it is straightforward to obtain the mixing parameters. 
These equations can also be applied directly to physical problems. An example,
neutrino oscillation in matter, was given in the previous section.

\section{Conclusion}

In this paper, we revisited the problem of the diagonalization of hermitian matrices.
While the process is known to be invariant under a number of symmetry operations, they are
usually not fully exploited. To remedy this, we propose to
parametrize a matrix $M$ by a set of variables $(\det M,d_{\alpha}^{(m)})$, where $d_{\alpha}^{(m)}$
is composed of the eigenvalues of its submatrices, $M_{\alpha}$.
Together with a similar set for the diagonal matrix $(M_{D})$, they satisfy a group of equations, Eq.(\ref{V}), 
which connects these variables with the mixing parameters, $W_{\alpha i}$.
These equations are very simple in structure, and are manifestly invariant under the symmetries mentioned above.
They can serve as the basic equations for diagonalization, since the parameters can be solved in terms of
each other without much difficulty. Another interesting feature of these equations is that, typically, each  
equation contains a subset of parameters pertaining to a certain problem.
This makes them more flexible in extracting informations from different physical situations. 
As an example, we considered neutrino oscillations in matter, where only $W_{ei}$ and $\lambda_{i}$ are involved.
Using the equations for $d_{e}$ and $t_{e}$, we found two new matter invariants $(I_{d}$ and $I_{t}$).
We verified that they are independent of matter effects, using available data which are inferred by
theoretical calculations. It is hoped that experimental results will be available soon so that
these equations can be directly tested with physical data.

\section*{Data Availability Statement} 
The data used to support the findings of this study are included within the article.

\section*{Conflicts of Interest} 
The authors declare no conflicts of interest.

\acknowledgments                 
One of us (SHC) is funded by Chang Gung Research Project, grant number: [BMRP834], 
of Chang Gung University, Taiwan. 
This manuscript has previously been presented in arXiv with the ID arXiv:2303.17087[hep-ph] 
at https://arxiv.org/abs/2303.17087.

\appendix
\section{Direct Proof of Eqs.(26-28)}

Eqs.(26-28), which connect the rephasing invariant parameters of $M$ with its 
eigenvalues and the mixing parameters, can also be derived directly.
To do this, we start from the explicit formula of diagonalization, given by
(with $[V]_{\alpha i}=V_{\alpha i}$, $[M_{D}]_{ii}=\lambda_{i}$):

\begin{eqnarray}\label{proof}
VM_{D}V^{\dag}&=&
\left(\begin{array}{ccc}
  M_{\alpha \alpha}& M_{\alpha \beta} & M_{\alpha \gamma} \\
  M_{\beta \alpha} & M_{\beta \beta} & M_{\beta \gamma} \\
  M_{\gamma \alpha} & M_{\gamma \beta} & M_{\gamma \gamma} \\
    \end{array}
    \right)  \nonumber \\
    &=&
    \left(\begin{array}{ccc}
   \Sigma W_{\alpha i}\lambda_{i} & \Sigma V_{\alpha i}V^{*}_{\beta i} \lambda_{i}& \Sigma V_{\alpha i}V^{*}_{\gamma i} \lambda_{i} \\
   \Sigma V_{\beta j}V^{*}_{\alpha j} \lambda_{j}& \Sigma W_{\beta j}\lambda_{j} &\Sigma V_{\beta j}V^{*}_{\gamma j} \lambda_{j}\\
   \Sigma V_{\gamma k}V^{*}_{\alpha k} \lambda_{k}& \Sigma V_{\gamma k}V^{*}_{\beta k} \lambda_{k} & \Sigma W_{\gamma k}\lambda_{k} \\
    \end{array}
    \right). 
\end{eqnarray}

From this we extract the rephasing invariant parameters of $M$, consisting of $\det M=\Pi \lambda_{i}$,
the diagonal elements $M_{\alpha \alpha}$ (or $t_{\alpha}=M_{\beta \beta}+M_{\gamma \gamma}$),
and the combinations $M_{\alpha \beta}M_{\beta \alpha}=|M_{\alpha \beta}|^{2}$
(or $d_{\gamma}=M_{\alpha \alpha}M_{\beta \beta}-|M_{\alpha \beta}|^{2}$).

From the diagonal elements, we find 
\begin{equation}
M_{\alpha \alpha}=\Sigma W_{\alpha i}\lambda_{i},
\end{equation}
which can be written in terms of the variables $t_{\alpha}=M_{\beta \beta}+M_{\gamma \gamma}$ and
$t_{i}=\lambda_{j}+\lambda_{k}$. We find
\begin{eqnarray}
t_{\alpha}&=& \Sigma (1-W_{\alpha i})\lambda_{i} \nonumber \\
&=& \Sigma'(W_{\alpha j}+W_{\alpha k})\lambda_{i} \nonumber \\
&=&\Sigma W_{\alpha i} t_{i},
\end{eqnarray}
which is Eq. (27).

We turn now to computing $d_{\gamma}$:
\begin{equation}\label{AI}
d_{\gamma}=(\Sigma W_{\alpha i}\lambda_{i})(\Sigma W_{\beta j}\lambda_{j})-
(\Sigma V_{\alpha i}V^{*}_{\beta i}\lambda_{i})(\Sigma V_{\beta j}V^{*}_{\alpha j}\lambda_{j}).
\end{equation}
The terms containing $\lambda^{2}_{i}$ cancel,
\begin{equation}
\Sigma W_{\alpha i}W_{\beta i}\lambda^{2}_{i}-
\Sigma (V_{\alpha i}V^{*}_{\beta i})(V_{\beta i}V^{*}_{\alpha i})\lambda^{2}_{i}=0,
\end{equation}
while terms proportional to $\lambda_{i}\lambda_{j}$, $i\neq j$, given by
\begin{eqnarray}
\Sigma_{i\neq j} \{[W_{\alpha i}W_{\beta j}+W_{\alpha j}W_{\beta i}]        
-[(V_{\alpha i}V^{*}_{\beta i})(V_{\beta j}V^{*}_{\alpha j})+(V_{\alpha j}V^{*}_{\beta j})(V_{\beta i}V^{*}_{\alpha i})]\}
\lambda_{i}\lambda_{j},   \nonumber
\end{eqnarray} 
can be reduced by using an identity [Eq. (30) in Ref.\cite{CK2016}],  
\begin{eqnarray}
[(V_{\alpha i}V^{*}_{\beta i})(V_{\beta j}V^{*}_{\alpha j})+c.c.]&=&2\Lambda_{\gamma k} \nonumber \\
&=&W_{\alpha i}W_{\beta j}+W_{\alpha j}W_{\beta i}-W_{\gamma k}.
\end{eqnarray}
Thus, Eq. (\ref{AI}) becomes
\begin{equation}
d_{\gamma}=\Sigma W_{\gamma k}d_{k}.
\end{equation}
Notice that Eqs. (26-27) can also be verified using numerical solutions. E.g., in Ref.~\cite{Denton1}, the matrix
\begin{eqnarray}\label{proof}
M&=&
\left(\begin{array}{ccc}
  1& 1 & -1 \\
 1& 3 & 1 \\
  -1 & 1 & 3 \\
    \end{array}
    \right) 
 \end{eqnarray}
was found to have the eigenvalues $(0,3,4)$ and 
\begin{eqnarray}\label{proof}
W_{\alpha i}&=&
\left(\begin{array}{ccc}
  2/3& 1/3 & 0 \\
 1/6& 1/3 & 1/2 \\
  1/6 & 1/3 & 1/2 \\
    \end{array}
    \right). 
 \end{eqnarray}
Also, from $M$ we can readily obtain $d_{\alpha}=(8,2,2)$, $t_{\alpha}=(6,4,4)$, which
verify Eq.(26-27).
Alternatively, given $\lambda_{i}$ and $(d_{\alpha},t_{\alpha})$, Eqs. (26-27) yield $W_{\alpha i}$,
without having to solve for the eigenvectors.

As another example, consider the matrix
\begin{eqnarray}
M&=&
\left(\begin{array}{ccc}
  1& \frac{1}{\sqrt{3}}e^{i\pi/6} & \frac{i}{\sqrt{3}} \\
 \frac{1}{\sqrt{3}}e^{-i\pi/6} & 1 & \frac{-1}{\sqrt{3}}e^{-i\pi/6} \\
  \frac{-i}{\sqrt{3}} & \frac{-1}{\sqrt{3}}e^{i\pi/6} & 1 \\
    \end{array}
    \right). 
 \end{eqnarray}
The usual procedure of solving the eigenvalue-eigenvector equations yields
\begin{equation}
\lambda_{i}=(0,1,2),
\end{equation}
\begin{equation}
V=\frac{e^{-i\pi/18}}{\sqrt{3}}
\left(\begin{array}{ccc}
  -\omega & 1 & 1 \\
 1 & -\omega & 1 \\
  1 & 1 & -\omega \\
    \end{array}
    \right),
     \end{equation}
 \begin{equation}
 \det V=+1, \thickspace \thickspace \omega=e^{i\pi/3}.
 \end{equation}

Alternatively, given $M$, we can readily find
$d_{\alpha}=2/3$, $t_{\alpha}=2$, $\alpha=(e,\mu,\tau)$.
Also, 
\begin{equation}
M_{e\mu}M_{\mu\tau}M_{\tau e}=\frac{i}{3\sqrt{3}},
\end{equation}
\begin{equation}
\det M=\Sigma M_{\alpha \alpha} d_{\alpha}-2\Pi \lambda_{\alpha}+2Re(M_{e\mu}M_{\mu \tau}M_{\tau e})=0.
\end{equation}
With $\Sigma \lambda_{i}=\frac{1}{2}\Sigma t_{\alpha}=3$, $\Sigma' \lambda_{i}\lambda_{j}=\Sigma d_{\alpha}=2$, and $\det M=0$, 
the eigenvalues follow from the characteristic equation, and $\lambda_{i}=(0,1,2)$.
The mixing parameters, from Eq.(29), are
\begin{equation}
W_{\alpha i}=\frac{(\lambda_{i}^{2}-t_{\alpha}\lambda_{i}+d_{\alpha})}
{(\lambda_{i}-\lambda_{j})(\lambda_{i}-\lambda_{k})}   
=\frac{1}{3},
\end{equation}
independent of $(\alpha,i)$.

Thus, Eqs. (26-28) provide a simple procedure to determine $(\lambda_{i},W_{\alpha i})$ from $M$.

\def\bibsection{\section*{Reference}} 

\end{document}